\documentclass{aastex62}

\usepackage{epsfig}
\renewcommand{\arraystretch}{1.2}
\newcommand{\hb}{H$\beta$}

\received{Month DD, 2020}
\revised{Month DD, 2020}
\accepted{Month DD, 2020}
\submitjournal{ApJ}
\defcitealias{Heasley+1974}{HMP}
\defcitealias{LabrosseGouttebroze2001}{LG}
\shorttitle{HELIUM CONTINUUM IN FLARE LOOPS}
\shortauthors{HEINZEL et al.}

\begin{document}
\title{SIGNATURES OF HELIUM CONTINUUM IN COOL FLARE LOOPS OBSERVED BY SDO/AIA}

\correspondingauthor{Petr Heinzel}
\email{pheinzel@asu.cas.cz}

\author[0000-0002-5778-2600]{PETR HEINZEL}
\affiliation{Astronomical Institute, The Czech Academy of Sciences, 25165 Ond\v{r}ejov, Czech Republic}

\author[0000-0001-5986-9948]{PAVOL SCHWARTZ} 
\affil{Astronomical Institute, Slovak Academy of Sciences, 05960 Tatransk\'{a} Lomnica, Slovakia}

\author[0000-0002-9690-8456]{JURAJ L\"{O}RIN\v{C}\'{I}K} 
\affiliation{Astronomical Institute, The Czech Academy of Sciences, 25165 Ond\v{r}ejov, Czech Republic}
\affiliation{Astronomical Institute, Faculty of Mathematics and Physics, Ke Karlovu 2027/3, 12116 Praha, Czech Republic}

\author[0000-0002-7444-7046]{J\'{U}LIUS KOZA} 
\affiliation{Astronomical Institute, Slovak Academy of Sciences, 05960 Tatransk\'{a} Lomnica, Slovakia}

\author[0000-0001-8489-4037]{SONJA JEJ\v{C}I\v{C}} 
\affiliation{Faculty of Education, University of Ljubljana, Kardeljeva plo\v{s}\v{c}ad 16, 1000 Ljubljana, Slovenia} 
\affiliation{Faculty of Mathematics and Physics, University of Ljubljana, Jadranska 19, 1000 Ljubljana, Slovenia}
\affiliation{Astronomical Institute, The Czech Academy of Sciences, 25165 Ond\v{r}ejov, Czech Republic}

\author[0000-0002-5778-2600]{DAVID KURIDZE} 
\affiliation{Department of Physics, Aberystwyth University, Ceredigion, SY23 3BZ, UK}

\begin{abstract}
We present an analysis of off-limb cool flare loops observed by SDO/AIA during the gradual phase of SOL2017-09-10T16:06 X8.2-class flare. 
In the EUV channels starting from the 335\,\AA\ one,  cool loops appear as dark structures against the bright loop arcade. These dark structures were
precisely coaligned (spatially and temporally) with loops observed by SST in emission lines of hydrogen and ionized calcium.
Recently published semi-empirical model of cool loops based on SST observations serves us to predict the level of hydrogen and helium
recombination continua. The continua were synthesized using an approximate non-LTE approach and theoretical spectra were then transformed
to AIA signals. Comparison with signals detected inside the dark loops shows that only in AIA  211\,\AA\ channel the computed level of
recombination continua is consistent with observations for some models, while in all other channels which are more distant from the continua edges the
synthetic continuum is far too low. In analogy with on-disk observations of flares we interpret the surplus emission as due to
numerous EUV lines emitted from hot but faint loops in front of the cool ones. 
Finally we briefly comment on failure of the standard absorption model when used for analysis of the dark-loop brightness.
\end{abstract}


\section{Introduction}

Visibility of dark prominence-like structures above the limb or against the disk
in otherwise 'hot' EUV channels like those of SOHO/EIT, TRACE or now SDO/AIA was interpreted
as the absorption of the background EUV line radiation from hot plasmas by cool prominences.
This absorption is due to photoionization of hydrogen and helium by EUV line photons. Within the
above channels, there is contribution of the hydrogen (Lyman continuum), below 504\,\AA\ the 
neutral helium is added and below 228\,\AA\ also ionized
helium contributes, but the helium continua dominate in the AIA channels \citep[see][]{AnzerHeinzel2005}.  
Many papers analyzed such dark prominences in order to derive their densities proportional to the amount
of absorbing material, for a review see \cite{Kucera2015}. Typical prominence densities derived
from such a diagnostics are in good agreement with those obtained from other analyses.
On the other hand, one could also consider the hydrogen and helium continuum {\em emission} as a
result of the photorecombinations which is the natural process at work. However, as shown by
\cite{Labrosse+2011}, Hinode/EIS prominence spectra do not 
show any detectable emission in the helium continua below 228\,\AA.
In typical prominence plasmas the electron densities are around $10^{10}$  $\mathrm{cm}^{-3}$
and this is probably too low for the recombinations to produce the observable continuum emission.
Although various studies dealt  with hydrogen and helium non\discretionary{-}{-}{-}LTE modeling in prominences \citep{Labrosse2015}, 
only the line intensities have been presented, no results for helium continua.

\begin{figure}
\begin{center}
\parbox{0.70\textwidth}{
\resizebox{0.70\textwidth}{!}{
\includegraphics{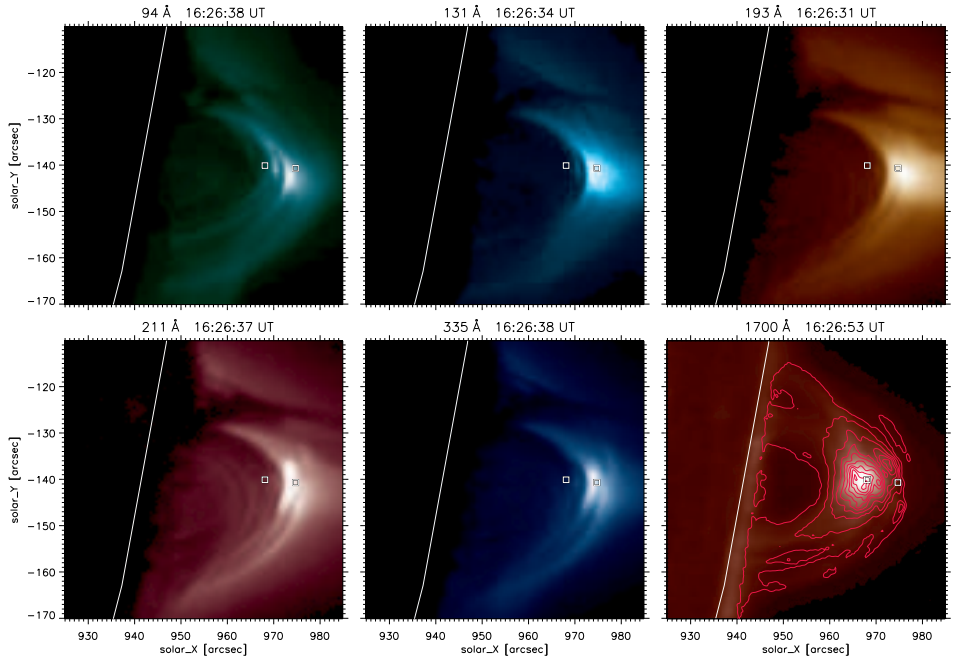}}}
\caption{Loop images in six AIA channels. Two areas (boxes) are marked in all images -- within dark loops, and within
bright loops just above them. Quiet corona was measured at the same height above the limb 
as the dark loop box, but far away from the active region. We show here also the image in 1700 \AA \,
coaligned with SST observations of cool loops (red contours).}
\end{center}
\label{fig:aia_imgs}
\end{figure}

Concerning cool flare loops, an appearance of dark loop structures against the bright EUV background was evidenced
both from off\discretionary{-}{-}{-}limb \citep{Jejcic+2018} as well as on-disk \citep{Song+2016} observations. 
As noticed in \cite{Jejcic+2018}, dark loops are clearly visible in SDO/AIA images of 10 September 2017
off\discretionary{-}{-}{-}limb flare loop system and we therefore focus in this paper on their detailed analysis. Our aim 
is to understand the nature of the dark-loop radiation, which could be, at least partially, due to helium
recombination continua. A similar analysis was carried out in \cite{Milligan+2012} and in \cite{Milligan+2013} for the case
of flare ribbons.

Off\discretionary{-}{-}{-}limb flare loops are routinely detected in emission in various spectral lines. Recently,
\cite{Koza+2019} used unique SST (Swedish Solar Telescope) observations of 10 September 2017 flare loops in the hydrogen H$\beta$ and
\ion{Ca}{2}\,8542\,\AA\ lines and performed the non\discretionary{-}{-}{-}LTE inversions in order to derive the physical parameters
of the cool loop plasma. They arrived at temperatures below $10^4$\,K, but of the great interest
are high electron densities around $10^{12}\,\mathrm{cm}^{-3}$, consistent with those obtained by \cite{Jejcic+2018} 
who analyzed the white\discretionary{-}{-}{-}light continuum emission from these loops as detected by SDO/HMI at 6173\,\AA.
We therefore use the model based on SST observations to predict the level of recombination continua below 504\,\AA\
(ionization edge of \ion{He}{1}) and to compare it with AIA observations in five channels, starting with 335\,\AA. 
We also briefly comment on the applicability of the absorption model for analysis of the dark-loop brightness.

\section{Extended flare loop system observed on 10 September 2017}

Right after the maximum of the SOL2017-09-10T16:06 X8.2-class flare, various dark loops could be well identified
on SDO/AIA images.
Here we focus on five AIA channels below the \ion{He}{1} continuum edge at 504\,\AA, where such dark absorbing loops 
were best visible (Figure~\ref{fig:aia_imgs}) and perform their photometric analysis. 
Observations made by the {\em Atmospheric Imaging Assembly} AIA \citep{Lemen+2012} instrument on board of the 
{\em Solar Dynamics Observatory} SDO \citep{Pesnell+2012} 
on September 10, 2017 at 16:26\,UT were selected because they are co\discretionary{-}{-}{-}temporal with 
SST imaging spectroscopy used for the non\discretionary{-}{-}{-}LTE diagnostics by \cite{Koza+2019}. 
In order to avoid a saturation, the observations in 94, 131, 193 and 211\,\AA\ channels were made in a flare mode with shorter exposures
$0.23$, $0.015$, $0.029$, and $0.397$\,s, respectively, compared to standard ones. 
However, flare-mode exposures as stored in the headers of the data fits files are sometimes unreliable and therefore we have 
calibrated them comparing normal and flare-mode exposures of the off-limb quiet corona. The resulting correction factors
to flare-mode exposures are 1.47, 2.28, 2.27, and 1.49 for these four AIA channels, respectively. 
In 335\,\AA\ channel the standard exposure of 2.9 s was used.
Observations in the 171\,\AA\ channel are not used because of overexposed pixels. 
The AIA data were reduced using the standard SolarSoft procedure \texttt{aia\_prep.pro}. After this basic reduction the data were corrected for internal instrumental scattering 
with the procedure \texttt{aia\_deconvolve\_richardsonlucy.pro} which deconvolves the point\discretionary{-}{-}{-}spread functions 
determined for all EUV channels by \cite{Poduval+2013} from the observed data. For the deconvolution itself the Richardson\discretionary{-}{-}{-}Lucy 
method was used \citep{Richardson1972,Lucy1974}. Uncertainties of the observed intensities were calculated using 
the \texttt{aia\_bp\_estimate\_error.pro routine}. In addition, we added to them 
28 \% calibration uncertainty of the instrument \citep{Boerner+2012}. 

In the AIA images shown in Figure~\ref{fig:aia_imgs} two box\discretionary{-}{-}{-}like areas were chosen, one at the top of the system of dark loops, and the second one 
inside a bright loop just above the apex of the dark\discretionary{-}{-}{-}loop system. The dark-loop box was coaligned with SST observations at about the
same time (see subsection 2.1 below). Average intensities in instrumental units $\mathrm{DN\,s}^{-1}\,\mathrm{pix}^{-1}$ were obtained from the two areas.
The so\discretionary{-}{-}{-}called foreground coronal intensity $I_{\mathrm{f}}$ is a contribution to the measured intensity from the corona in front of the loop arcade. 
Taking into account 
the fact that geometrical thickness of the loops is negligible as compared to very extended solar corona, the foreground intensity can be expressed as one half of the intensity 
of the quiescent corona $I_{\mathrm{QS}}$ measured at the same height above the limb as the dark\discretionary{-}{-}{-}loop intensity but far enough from the active
region. 
Intensities $I_{\mathrm{DL}}$ (dark loop), $I_{\mathrm{BL}}$ (bright loop above), and $I_{\mathrm{QS}}$ measured in corresponding boxes in the five AIA channels
are shown in Table~\ref{table:tauaiacalc}.

As a next step we use the model obtained by \cite{Koza+2019} from the non\discretionary{-}{-}{-}LTE inversions of
lines observed by SST to derive the radiation properties of the cool loops in AIA bands. The densities from the  
best-fit SST model lead to total optical thicknesses shown in Table~\ref{table:models} (for their evaluation see the next section). 
Such very large optical thicknesses clearly indicate that all background EUV radiation, i.e. along the line of sight behind the cool loops, must be totally absorbed.
The question then arises {\em What is the nature of the AIA signal within the cool-loop box ?} 
A natural possibility is that the
detected emission in dark loops is the radiation of cool loops themselves in the helium
recombination continua. This would not be very surprising because of high densities
found independently in these loops. For example, in solar prominences having almost two orders of magnitude lower densities than those
found in 10 September 2017 cool loops, 
the resonance continuum below 228\,\AA\ was not detected by \cite{Labrosse+2011} who analyzed the Hinode/EIS spectra,
although these authors mention some other space observations where such a continuum was present. We therefore perform the
theoretical non-LTE synthesis of such continua below 504\,\AA\ and compare their intensities with detected AIA signals in all
considered channels.

\subsection{Spatial alignment of SDO/AIA and SST observations} 

In order to be able to analyze our AIA observations in terms of SST models, we need a careful spatial alignment of AIA
and SST images. To co\discretionary{-}{-}{-}align the space- and ground\discretionary{-}{-}{-}based imagery of the
flare loops, the SST/CHROMIS \hb\ image in wavelength\discretionary{-}{-}{-}integrated
intensities is created \citep[see the left panel of Figure~2 in][]{Koza+2019},
allowing clearly recognizable common features to be identified in 
the AIA 1700\,\AA\ passband where the cool lines dominate \citep{Jejcic+2018}.
For the co\discretionary{-}{-}{-}alignment the function
\texttt{auto\_align\_images.pro} is used, which is
implemented within the IDL SolarSoft System
\citep{FreelandHandy1998}. Through the cross\discretionary{-}{-}{-}correlation, a
satisfactory spatial alignment of co-temporal AIA and SST images is
achieved as shown in Figure~\ref{fig:aia_imgs}. 

\begin{table}
\centering{
\caption{Measured intensities $I_{\mathrm{DL}}$, $I_{\mathrm{BL}}$ and $I_{\mathrm{QS}}$ in the five AIA channels, with uncertainties less than 30 \%.
Positions of the two areas are as follows:
Dark\discretionary{-}{-}{-}loop area (DL) is located at the position solar\_X=$+968$\,arcsec\ solar\_Y=$-140$\,arcsec,
bright\discretionary{-}{-}{-}loop area (BL) is located just above the dark loops at the position
solar\_X=$+975$\,arcsec\ solar\_Y=$-141$\,arcsec, and
bright\discretionary{-}{-}{-}loop areas (BL1) were taken at the same height as the DL box (we use the interpolated value
from both sides around the dark loops).
The quiet\discretionary{-}{-}{-}corona areas (QS) were taken sufficiently far from the active region (out of the frame of our AIA images).
}
\label{table:tauaiacalc}
\begin{tabular}{ccccc}
\hline \\[-2.8ex]
AIA channel & $I_{\mathrm{DL}}$ & $I_{\mathrm{BL}} $ & $I_{\mathrm{BL1}} $ & $I_{\mathrm{QS}}$  \\
$\left[\mathrm{\AA}\right]$  &  $[\mathrm{DN\,s}^{-1}\,\mathrm{pix}^{-1}]$  & $[\mathrm{DN\,s}^{-1}\,\mathrm{pix}^{-1}]$ & $[\mathrm{DN\,s}^{-1}\,\mathrm{pix}^{-1}]$  & 
$[\mathrm{DN\,s}^{-1}\,\mathrm{pix}^{-1}]$ \\[1mm]
\hline \\[-2.8ex]
94 & 10000 & 67000 & 20000  & 250  \\
131 & 13000 & 639000 & 150000 & 1100 \\
193 & 39000 & 1116000 & 180000 & 1600 \\ 
211 & 5800 & 70000 & 20000 & 300 \\
335 & 100 & 1500 & 1100 & $ 5 $ \\

\hline
\end{tabular}
}
\end{table}

\section{Synthetic continuum intensity}

So far no helium continuum intensities have been computed for prominence\discretionary{-}{-}{-}like structures including cool flare loops.
\cite{Heasley+1974} (hereafter referred to as \citetalias{Heasley+1974}) and
\cite{LabrosseGouttebroze2001}
presented the results of non-LTE helium line formation under prominence conditions but they did not synthesize the helium resonance continua.
At the wavelengths of interest below 504\,\AA, three resonance continua are considered: the hydrogen Lyman continuum (head at 912\,\AA),
\ion{He}{1} continuum (head at 504\,\AA) and \ion{He}{2} continuum (head at 228\,\AA).
The continuum absorption coefficient takes the form \citep{AnzerHeinzel2005, HubenyMihalas2015}

\begin{equation}
  \kappa_{\nu} = \sigma_{\mathrm{H}\,{\normalfont\textsc{i}}}(\nu)\,n_{\mathrm{H}\,{\normalfont\textsc{i}}} + \sigma_{\mathrm{He}\,{\normalfont\textsc{i}}}(\nu)\,
  n_{\mathrm{He}\,{\normalfont\textsc{i}}} + \sigma_{\mathrm{He}\,{\normalfont\textsc{ii}}}(\nu)\,n_{\mathrm{He}\,{\normalfont\textsc{ii}}}\,,
\label{eq:kappa_nu}
\end{equation} 
where $n$ is the ground-state population of the respective ion and $\sigma(\nu)$ is the frequency-dependent absorption cross-section
for photoionization ($\sigma_{\mathrm{He}\,{\normalfont\textsc{ii}}}(\nu)$ is effective below 228\,\AA) . 
To compute $\sigma$ we use standard formulas for hydrogen and \ion{He}{2} \citep{AnzerHeinzel2005} and the polynomial
expansion of \cite{Rumph+1994} for \ion{He}{1}.
The correction for induced emission is negligible in EUV. The ground states of the
respective ions, from which the photoionizations take place, have dominant populations and thus we can replace them 
with total populations of the ions. 
Adding the hydrogen continuum opacity computed consistently with the MALI hydrogen non\discretionary{-}{-}{-}LTE code \citep{Heinzel1995, Koza+2019}, 
we get the continuum optical thicknesses presented in Table~\ref{table:models}, where $\tau_{228}$ is in good agreement with \citetalias{Heasley+1974}
for their model HMP7. 
An interesting feature here is that the
total opacity of helium continua below 228\,\AA\ practically does not depend on the ionization structure of helium - this is because both helium cross-sections
are almost equal close to the continuum head. Then the helium contribution to total $\tau$ is proportional only to hydrogen column density (see also section 5).

The emission coefficient can be expressed according to \cite{HubenyMihalas2015} as

\begin{equation}
  \eta_{\nu} = \left[\hspace{0.2ex}\sigma_{\mathrm{H}\,{\normalfont\textsc{i}}}(\nu)\,n^*_{\mathrm{H}\,{\normalfont\textsc{i}}} + \sigma_{\mathrm{He}\,{\normalfont\textsc{i}}}(\nu)\,
    n^*_{\mathrm{He}\,{\normalfont\textsc{i}}} + \sigma_{\mathrm{He}\,{\normalfont\textsc{ii}}}(\nu)\,n^*_{\mathrm{He}\,{\normalfont\textsc{ii}}}\hspace{0.2ex}\right]\,B_{\nu}(T)\,,
\label{eq:etamu}
\end{equation}
where we also can neglect the term $(1 - e^{-h\nu/kT})$. Here $n_i^*=n_{i+1} n_{\rm e} \Phi_i(T)$ are the LTE populations of the ground states of respective ions
computed for {\em actual} electron densities $n_\mathrm{e}$ and non\discretionary{-}{-}{-}LTE populations $n_{i+1}$ of higher ions, $\Phi_i(T)$ is the Saha-Boltzmann
factor and $B_{\nu}(T)$ the Planck function. Again the last term is effective only below 228\,\AA. The continuum source function is then $S_{\nu} = \eta_{\nu} / \kappa_{\nu}$, proportional to
$B_{\nu}(T)$ and to $n_{\rm e}$. 
For a given model we compute the electron density (here equal to proton density) and the non-LTE hydrogen ground-state population using
the hydrogen code MALI \citep{Heinzel1995}, with partial frequency redistribution in the Lyman lines.
The non-LTE populations of \ion{He}{1}, \ion{He}{2}, and \ion{He}{3} ground-states (here $n_1$, $n_2$ and $n_3$, respectively) 
are computed following the approach of \cite{Avrett+1976}. We write the ionization equilibrium
equations for these three helium ions and get the population ratios 

\begin{eqnarray}
a=\frac{n_2}{n_1} & = & \frac{R_{12} + C_{12}}{R_{21}} \nonumber \\
b=\frac{n_3}{n_2} & = & \frac{R_{23} + C_{23}}{R_{32}} \, ,
\end{eqnarray}
where $R_{12}$ and $R_{23}$ are the photoionization rates, $C_{12}$ and $C_{23}$ the collisional ionization rates according to \cite{MihalasStone1968} and
\cite{Avrett+1976}, respectively, and $R_{21}$ together with $R_{32}$ are the
radiative recombination rates to all levels of the respective ion 
(we neglect dielectronic and collisional three-body recombinations). To evaluate the photoionization rates we use the above-mentioned
cross sections and the external radiation field illuminating the cool loop. The latter is taken from HMP, but was modified below 228\,\AA\ where we use the
EUV spectral measurements of the quiet Sun from OSO-7 satellite reported by \cite{Linsky+1976}. All these incident intensities are for quiet-Sun chromospheric
and coronal illumination. In order to account for enhanced illumination of the cool loops by surrounding hot EUV loops, we enhance the whole considered
spectrum below 504\,\AA\ by a certain factor obtained from our AIA measurements (Table 1). A lower limit is the radiation detected within 
the dark-loop box, i.e. we assume that
the cool SST loop structures can be illuminated by hot loops emitting in front of them (see also the discussion in next sections). An upper limit is the
radiation from hot loops located around the cool ones. We compute the enhancement factors as the ratios $I_{\rm{DL}}/(I_{\rm{QS}}/2)$ and $I_{\rm{BL}}/(I_{\rm{QS}}/2)$
which, using data from Table 1, gives averaged values 43 (DL models) and 820 (BL models), respectively. Contrary to \cite{Avrett+1976}, we ignore here the fact that the incident
radiation is partially absorbed at a given atmospheric depth from the surface - our simulations have shown that this effect is of secondary importance in our
approximate modeling. As a test case we considered the prominence model HMP7, illuminated by the quiet-Sun radiation. The resulting continuum intensity
at 228\,\AA\ is 3.6 $\times 10^{-13}$ which fits quite well in the range of OSO-7 prominence observations reported by \cite{Linsky+1976}. 
In present analysis we do not solve the full transfer problem for helium continuum but observing that
$\tau$ in Table~\ref{table:models} is very large for all considered models, we use the Eddington-Barbier relation
\citep{HubenyMihalas2015} to estimate the emergent continuum intensity, i.e. $I_{\nu} \simeq S_{\nu}$, where the source function
is determined at {\em surface layers} where $\tau_{\nu} \simeq 1$. This approximation thus works best close to the continuum limit (head), at deeper layers the source
function may differ. 

In our simulations we consider six loop models (Table~\ref{table:models}), 
based on the best\discretionary{-}{-}{-}fit inversion model obtained from the SST spectra \citep[Table~3 in][]{Koza+2019}.
They are isothermal and isobaric having quasi-constant hydrogen density through the loop thickness $D$.
Surface electron densities $n_{\rm e}$ are used to compute the synthetic continuum. In our hydrogen code the electron density is equal
to proton density and thus the ratio $n_{\rm e}/n_{\rm H}$ is the hydrogen ionization degree. We use the original SST cool-loop model and two of its
variants with higher temperatures of 15000 and 20000 K, simulating cool loops
having a transition region to the corona (like PCTR in prominences). Each such model has been computed using two limiting enhanced illumination
factors. In Table~\ref{table:models} we show the helium ionization ratios $a$ and $b$ and present the synthetic cgs intensities at the head of \ion{He}{2} continuum at
228\,\AA\ and at 211\,\AA, two wavelengths representative of the OSO-7 spectra and AIA channel together with Hinode/EIS, respectively.
Note a decrease of intensities with increasing temperature at 228\,\AA, but an increase at 211 - this is due to behaviour of the helium
continuum source function as shown in \cite{Avrett+1976}.
Optical thickness is very large at all considered wavelengths as shown in Table~\ref{table:models}. At 335\,\AA\, it represents a lower limit because we 
do not consider a restricted penetration of the incident EUV radiation into the loop (this is not a problem for wavelengths below 228\,\AA\, where $\tau$ is practically
insensitive to helium ionization). Note also that the hydrogen Lyman-continuum opacity contributes significantly at low temperatures below 10000 K.
The whole synthetic spectrum is then used to obtain AIA signals in all bands as described in the next section. 

\section{Comparison with AIA observations} 

AIA intensities are in units of $\mathrm{DN\,s}^{-1}\,\mathrm{pix}^{-1}$ which cannot be directly compared with synthetic intensities
expressed in $\mathrm{erg}\,\mathrm{s}^{-1}\mathrm{cm}^{-2}\,\mathrm{sr}^{-1}\mathrm{Hz}^{-1}$ (called cgs units for brevity). This is because the AIA data
can hardly be calibrated to these absolute energy units. We therefore proceeded in an opposite way and converted the synthetic
intensities in cgs units to synthetic  $\mathrm{DN\,s}^{-1}\,\mathrm{pix}^{-1}$ signals which are readily comparable with observations.
To do so, we multiplied the modelled spectra with the response functions of selected filter channels, which we corrected for the sensitivity decay, 
and with a constant standing for the conversion between the solid angle and AIA pixel. The resulting products were then integrated over the bandpasses of the individual filter channels. 
Note that the response functions of EUV filter channels are typically broad and those of the 193 and 211 channels cover wavelengths longer than the \ion{He}{2} continuum head at
228\,\AA. Therefore, the spectra were only integrated in wavelength ranges corresponding to $99$\,\% of the total observed signal.

\begin{figure} 
\begin{center}
\includegraphics[width=8.0cm]{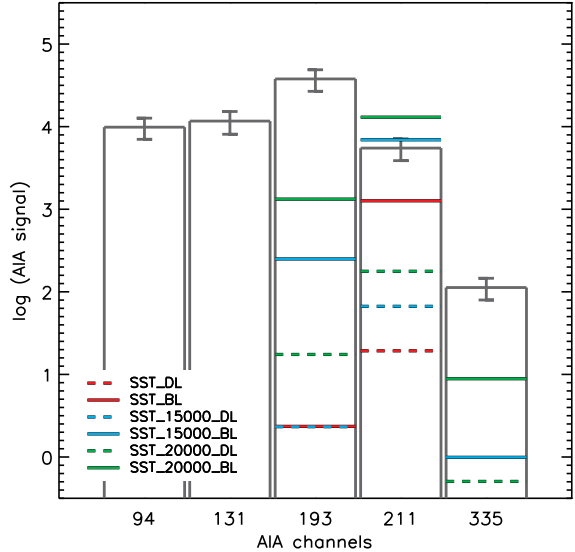} 
\end{center}
\caption{Comparison of the observed and synthetic signals in AIA channels (in $\mathrm{DN\,s}^{-1}\,\mathrm{pix}^{-1}$ units). 
Black columns represent the observed signal with corresponding error bars, the synthetic signals from different models are shown as color lines. In two leftmost channels
the synthetic signal is too low to be displayed.}
\label{fig_synth_obs_aia}
\end{figure}

Results of the conversion of synthetic spectra into AIA signals are shown in Table~\ref{table:models} and in Figure~\ref{fig_synth_obs_aia} where the synthetic observables
are indicated by color lines. Dashed lines show the lower limit of the AIA signal for the case of DL illumination, while the full color lines represent the upper limit
for BL illumination. Note that these signals are dominated by respective helium recombination continua, while the hydrogen Lyman continuum is negligible.
Black columns show levels of the observed intensities from Table~\ref{table:tauaiacalc}, together with their error bars (after the foreground subtraction).
Fast decrease of the synthetic continuum intensity with decreasing wavelength (compare channels 211 and 193) is related to behaviour of the Planck function in far EUV.

We see that the SST model with BL illumination gives the intensity which is about 20 \% of the observed one in AIA  211\,\AA\ channel, 
while the same model with DL  illumination has much
lower intensity. Comparison in this channel is largely improved if we increase the loop temperature to 15000 or even 20000 K, but other channels below 211\,\AA\ have
intensities still far below the observed ones. We discuss this behaviour in the last section. Since we deal here with the surface layers where the 211\,\AA\  continuum
is formed (due to its large opacity), the enhanced temperature can be representative of a PCTR while the central parts of the loop, if cooler, can still be consistent
with the SST model. Note that the SST model is based on lines having a moderate optical thickness and thus being formed deeper, out of PCTR.
In Table~\ref{table:models} and Figure~\ref{fig_synth_obs_aia} we also show
the situation in the 335\,\AA\ channel, again for the best-fit SST model. The synthetic intensity is much lower in comparison with the AIA signal, similarly as in other AIA channels
below 200\,\AA. The 335\,\AA\ channel is already far from the 504\,\AA\ continuum head of neutral helium and thus it is not surprising that the intensity substantially drops.
But we see an indication of the 228\,\AA\ jump between channels 211 and 335. The reason why the synthetic signals are much lower compared to the observed ones, 
except for
211\,\AA\ channel, is discussed in the last section.

\begin{table}
\centering{
\caption{Grid of six cool\discretionary{-}{-}{-}loop  SST models used to simulate the AIA signals. The models are isothermal-isobaric with a uniform gas pressure
$p$=9.1 dyn cm$^{-2}$ and microturbulent velocity $v_{\rm t}$ = 24 km s$^{-1}$. The thickness of SST models is $D$=5100 km. HMP7 is the prominence model
of \cite{Heasley+1974} shown here for reference, with $p$=0.02 dyn cm$^{-2}$, $v_{\rm t}$ = 5 km s$^{-1}$ and $D$=6000 km. For other quantities see the text.\\
cgs = erg s$^{-1}$ cm$^{-2}$ sr$^{-1}$ Hz$^{-1}$, AIA = $[\mathrm{DN\,s}^{-1}\,\mathrm{pix}^{-1}]$
}
\label{table:models}
\begin{tabular}{ccccccccccccccc}
\hline \\[-2.8ex]
  model & $T$ & $n_{\mathrm{H}}$ & $n_{\mathrm{e}}$ & $a$ & $b$ & $I_{228}$ & $I_{211}$ & $I_{335}$ &
  $I_{211}$ & $I_{193}$ & $\tau_{335}$ & $\tau_{228}$ & $\tau_{211}$ & $\tau_{193}$ \\
  & [kK] & $\left[\mathrm{cm}^{-3}\right]$ & $\left[\mathrm{cm}^{-3}\right]$ & & &  [cgs] & 
  [cgs] & [AIA] & [AIA] & [AIA] & & & & \\
\hline \\[-2.8ex]
SST-DL & 8.7 & 6.23+12 & 7.22+11 & 3.4-1 & 3.2-3 & 6.1-12 & 2.2-14 & 4.0-5 & 19.3 & 3.5-2 & 1208 & 641 & 536 & 436\\ 
SST-BL & 8.7 & 6.23+12 & 7.22+11 & 6.6 & 6.0-2 & 4.0-10 & 1.5-12 & 2.0-4 & 1264.7 & 2.3 & 513 & 596 & 486 & 383\\
SST-DL-15 & 15 & 2.10+12 & 2.08+12 & 1.8-1 & 1.6-3 & 3.7-12 & 1.50-13 & 5.4-2 & 66.7 & 2.3 & 323 & 178 & 151 & 125\\
SST-BL-15 & 15 & 2.10+12 & 2.08+12 & 3.4 & 3.0-2 & 3.7-10 & 1.6-11 & 9.9-1 & 6926.7 & 250.1 & 85 & 167 & 138 & 109\\
SST-DL-20 & 20 & 1.57+12 & 1.56+12 & 3.1-1 & 2.6-3 & 4.5-12 & 4.3-13 & 5.1-1 & 177.3 & 17.4 & 133 & 132 &112 & 92\\
SST-BL-20 & 20 & 1.57+12 & 1.56+12 & 5.6 & 4.9-2 & 3.2-10 & 3.2-11 &  8.9 & 13017.0 & 1325.9 & 41 & 122 & 100 & 80\\
HMP7 &  8 & 1.00+10 & 7.10+9 & 7.7-1 & 7.1-3 & 3.6-13 & 7.7-16 &  3.6-7 & 8.6-1 & 8.3-4 & 1.8 & 1.2 & 1.0 & 0.8 \\
\hline
\end{tabular}
}
\end{table}

\section{Comments on standard absorption model}
 
In a direct analogy with off-limb prominences one is tempted to apply a standard absorption model to derive the optical thickness of
the recombination continua and from that to estimate the density.
Assuming that a system of bright loops is located behind the dark ones and therefore their intensity measured just above the dark loops 
is representative of the background EUV radiation (Figure 1 and Table 1), the optical thickness can be calculated as
\begin{equation}
\tau=-\ln\left(\frac{I_{\mathrm{DL}}-I_{\mathrm{f}}}{I_{\mathrm{BL}}}\right)\,,
\label{eq:tau_from_aia}
\end{equation}
where $I_{\mathrm{f}}=I_{\mathrm{QS}}/2$. Resulting values of $\tau$ are shown in Table 3.
These values indicate a partial absorption of the
background EUV radiation, a standard scenario considered for interpreting dark prominences. Using channels 211 and 193, one can estimate the hydrogen column density $N_{\rm H}$
in cool loops  using the relation \citep{AnzerHeinzel2005} 

\begin{equation}
N_{\mathrm{H}} = \tau / \, 0.1 \sigma_{\mathrm{He}\,{\normalfont\textsc{i}}} \,,
\label{eq:nhdef}
\end{equation} 
where $\sigma_{\mathrm{He}\,{\normalfont\textsc{i}}}$ is the photoionization cross\discretionary{-}{-}{-}section of \ion{He}{1} (\cite{Rumph+1994} 
and $0.1$ is the helium abundance relative to hydrogen. At low temperatures we can neglect
the \ion{He}{3} ion in estimating total helium density. In this relation we also neglected the hydrogen Lyman\discretionary{-}{-}{-}continuum opacity and used to
advantage the fact that $\sigma_{\mathrm{He}\,\normalfont\textsc{i}} \simeq \sigma_{\mathrm{He}\,\normalfont\textsc{ii}}$ below 228\,\AA. Resulting hydrogen 
column densities are summarized
in Table~\ref{table:absmodel} for two AIA channels where the latter approximation holds best. However, this result looks very surprising. 
The obtained column density,
divided by a characteristic thickness of 5000 km (Table 2), is about
two orders of magnitude lower than the hydrogen densities derived in the {\em same loops} from other diagnostics like optical-line inversions
\citep{Koza+2019} or from white\discretionary{-}{-}{-}light emission detected by SDO/HMI \citep{Jejcic+2018}.
Such density will be even lower if we add the hydrogen Lyman-continuum opacity to evaluation of
total $\tau$, or, if we use less bright hot loops with $I_{\mathrm{BL1}}$ as the background radiation.
The reason for such a large discrepancy lies in improper application
of the absorption model. While in case of prominences only the absorption of the background radiation plays a major role, in cool flare loops we
detect an extra emission $I_{\mathrm{DL}}-I_{\mathrm{f}}$ which, as demonstrated in this study, is not a partially absorbed background radiation because
we know from independent diagnostics that the opacity is very large.
Low values of $\tau$ obtained from the absorption model are thus an artefact of ignoring emission sources such as the continuum radiation discussed in this
paper or EUV line emission by weak hot loops in front of the cool ones. 

As discussed by \cite{AnzerHeinzel2005}, the so-called {\em emissivity deficit} (or formerly called a volume blocking) can play a 
role in the absorption model in case of prominence structures significantly extended along the line of sight.
This may generally apply also to
an arcade of flare loops seen off-limb, but a detailed geometry is difficult to assess due to extreme complexity of the coronal emission. 
However, an emissivity deficit would play no role in case of our dark loops
which are so opaque that the background radiation is totally absorbed. It may only have some effect on determination of the foreground coronal intensity
$I_{\rm f}$, but this is rather negligible.

\begin{table}
\centering{
\caption{
Empirical optical thickness $\tau$ computed with the standard absorption model and using $I_{\mathrm{BL}}$ as the background (the accuracy is
better than 20 \%). $N_{\rm H}$ is the
column density derived from channels 193 and 211 (see the text).}
\label{table:absmodel}
\begin{tabular}{cccc}
\hline \\[-2.8ex]
AIA channel & $\tau$ & $\sigma_{\mathrm{He}\,{\normalfont\textsc{i}}}$ & $N_{\mathrm{H}}$  \\
$\left[\mathrm{\AA}\right]$  &  & $\left[\mathrm{cm}^{2}\right]$ & $\left[\mathrm{cm}^{-2}\right]$  \\[1mm]
\hline \\[-2.8ex]
94 & 1.9 & $-$ & $-$ \\
131 & 4.0 & $-$ & $-$ \\
193 & 3.4 & 1.19$-$18 & 3.0+19 \\ 
211 & 2.5 & 1.44$-$18 & 1.9+19 \\
335 & 2.6 &  $-$ & $-$ \\
\hline
\end{tabular}
}
\end{table}

\section{Discussion and conclusions}

Our analysis of SDO/AIA cool loops seen in absorption shows that in case of 
dense cool flare loops the standard absorption model
leads to unrealistic underestimation of the opacity (optical thickness) and thus the plasma density, by almost two orders of
magnitude. A large opacity was derived from the best\discretionary{-}{-}{-}fit inversion model of the same loops obtained by \cite{Koza+2019}.
This means that the background radiation is totally absorbed. To understand the nature of the radiation detected in dark AIA loops, we have tested
six loop models and estimated the level of the helium recombination continuum in all channels. While in channel 211\,\AA\ the
observed signal can be represented by some models, in other channels the synthetic observables
are much lower compared to the AIA signal. Our interpretation is that while the 211\,\AA\ channel may be dominated by the helium
continuum in some cases, other channels show the emission which seems to be due to weak hot loops projected against the dark ones -
note that the whole loop arcade is rotated with respect to the line of sight \citep{Kuridze+2019}.
This is similar to observations of on-disk flares \citep{Milligan+2012, Milligan+2013}, where the ribbons produce 
the helium continuum which is visible close to continuum ionization edges at 504\,\AA\ and 228\,\AA, but further from them strong EUV lines
and free-free continuum dominate.
There exists also a possibility that the helium continuum itself is enhanced in a foreground hotter (transition-region) plasma
because it is strongly dependent on plasma temperature - we see such a transition region in 1700\,\AA\ channel.
We also compared our model results with the AIA signal in 335\,\AA\ band and the result is similar to other channels below 200\,\AA .
The  \ion{He}{1} recombination continuum seems to be again dominated by other EUV emissions.
Our analysis thus suggests that we may detect the helium recombination
continuum only in the 211\,\AA\ band. 

The fact that we detect certain amount of radiation in otherwise totally absorbing cool loops affects the standard absorption model,
leading to spurious values of $\tau$. Such too low values are the artefact  of ignoring such emission in the standard model and thus this
model is inapplicable to cool flare loops with high density, contrary to the case of quiescent prominences.
This might also be a reason why the empirical $\tau$ in Table~\ref{table:tauaiacalc} does not decrease with decreasing
wavelength as expected. 

As the reader can see, the quantitative analysis using the SDO/AIA data is rather cumbersome due to calibration issues and also due
to inability to distinguish between the continuum and EUV line contributions to the AIA signals. In the future work we will explore the possibility of using spectra 
of these loops taken by Hinode/EIS. In Table 2 we show the intensities at 211\,\AA\ which is around the limit of shorter-wavelength band of EIS where
\cite{Young+2007} detected helium continuum in an active region. We will also make a more detailed non-LTE diagnostics, focusing
on the helium continuum
formation with a complex behaviour of EUV photoionization by the surrounding arcade of hot flaring loops and using the
up-to-date helium atomic data.

\acknowledgements
{The AIA data used are courtesy of SDO (NASA)
and the AIA consortium. The
authors acknowledge support from the grant 19-17102S of the Czech
Funding Agency. P.S. and J.K. acknowledge the project VEGA
2/0048/20.
S.J. acknowledges the support from the
Slovenian Research Agency No. P1-0188.
The Swedish 1-m Solar Telescope is operated by the Institute for Solar Physics of Stockholm University
and is located on the island of La Palma at the
Spanish Observatorio del Roque de los Muchachos (Instituto de
Astrofisica de Canarias). The Institute for Solar Physics is supported
by a grant for research infrastructures of national importance from
the Swedish Research Council (No. 2017-00625).
D.K.  has received funding from the S\^{e}r Cymru II Part-funded by the
European Regional Development Fund through the Welsh Government
and from the Georgian Shota Rustaveli National
Science Foundation project FR17 323. The authors also acknowledge the support by project RVO:67985815 of the
Astronomical Institute of the Czech Academy of Sciences. The authors thank Dr. J. Dud\'\i k for useful discussions
and comments and the anonymous referee for helpful advice. 
}

\facility{SDO(AIA), SST(CHROMIS)}


\end{document}